\documentclass[journal=jacsat,manuscript=article,layout=twocolumn]{achemso}

\usepackage{siunitx} 
\usepackage{tikz}
\usepackage{comment}
\usepackage{tikz-3dplot}
\usepackage{tikzscale}
\usepackage{pgfplots}
\usepackage[version=3]{mhchem}
\usepackage{physics}
\usepackage{breqn}

\renewcommand{\vec}[1]{\mathbf{#1}}
\newcommand{\rmi}{\mathrm{i}}
\newcommand{\rme}{\mathrm{e}}
\newcommand{\rmd}{\,\mathrm{d}}
\author{Jussi Kelavuori}
\affiliation[TAU]
{Photonics Laboratory, Physics Unit, Tampere University, FI-33014 Tampere, Finland}
\author{Viatcheslav Vanyukov} 
\affiliation[UEF]
{Faculty of Science and Forestry, Department of Physics and Mathematics, University of Eastern Finland, FI-80101 Joensuu, Finland}
\author{Timo Stolt}
\affiliation[TAU]
{Photonics Laboratory, Physics Unit, Tampere University, FI-33014 Tampere, Finland}
\author{Petri Karvinen}
\author{Heikki Rekola}
\affiliation[UEF]
{Faculty of Science and Forestry, Department of Physics and Mathematics, University of Eastern Finland, FI-80101 Joensuu, Finland}
\affiliation[TAU]
{Photonics Laboratory, Physics Unit, Tampere University, FI-33014 Tampere, Finland}
\author{Tommi K. Hakala}
\affiliation[UEF]
{Faculty of Science and Forestry, Department of Physics and Mathematics, University of Eastern Finland, FI-80101 Joensuu, Finland}
\author{Mikko J. Huttunen}
\email{mikko.huttunen@tuni.fi}
\affiliation[TAU]
{Photonics Laboratory, Physics Unit, Tampere University, FI-33014 Tampere, Finland}

\title[Thermal Control of Plasmonic Surface Lattice Resonances]
  {Thermal Control of Plasmonic Surface Lattice Resonances}

\abbreviations{IR,NMR,UV}
\keywords{metamaterials, plasmoncs, surface lattice resonance}

\let\oldmaketitle\maketitle
\let\maketitle\relax

\begin{document}
\twocolumn[
\begin{@twocolumnfalse}
\oldmaketitle
\begin{abstract}
Plasmonic metasurfaces exhibiting collective responses known as surface lattice resonances (SLRs) show potential for realizing tunable and flat photonic components for wavelength-selective processes, including lasing and optical nonlinearities. However, post-fabrication tuning of SLRs remains challenging, limiting the applicability of SLR-based components. 
Here, we demonstrate how the properties of high quality factor SLRs are easily modified by breaking the symmetry of the nanoparticle surroundings.
We break the symmetry by changing the refractive index of the overlying immersion oil simply by controlling the ambient temperature of the device. We show that already modest temperature changes of \SI{10}{\celsius} can increase the quality factor of the investigated SLR from 400 to 750.
Our results demonstrate accurate and reversible modification of the properties of the SLRs, paving the way towards tunable SLR-based photonic devices. On a more general level, our results demonstrate how symmetry breaking of the surrounding dielectric environment can be utilized for efficient and potentially ultrafast modification of the SLR properties.
\end{abstract}
\end{@twocolumnfalse}
]
Optical metamaterials are artificial structures that allow control of light in ways not found in nature\cite{Soukoulis2011}. Current research on optical metamaterials covers a wide spectrum of topics including saturable absorption~\cite{Dayal2013}, nanoscale phase-engineering~\cite{Hu2019}, epsilon-near-zero behavior~\cite{Alu2007,Alam2016}, and supercontinuum generation~\cite{Krasavin2016,Chen2018}. Furthermore, considerable interest has been focused on plasmonic metasurfaces consisting of metallic nanoparticles (NPs). Metallic NPs exhibit collective responses of conduction electrons known as localized surface plasmon resonances (LSPRs)~\cite{Maier2007}. On top of their high modifiability, the LSPRs increase the local near fields at the NP surface enhancing the occurring light--matter interaction, making applications such as biosensing~\cite{Unser2015}, lasing~\cite{Pourjamal2019} and nonlinear optical processes~\cite{Kauranen2012review} possible. Unfortunately, metallic NPs suffer from high losses due to ohmic nature of metals. The losses can be reduced by arranging NPs periodically and utilizing collective responses of periodic structures known as surface lattice resonances (SLRs)~\cite{Kravets2018}. SLRs are diffractive--plasmonic hybrid resonances associated with high quality factors ($Q$-factors) that can reach values above 2000 \cite{Bin-Alam2021b}. Consequently, SLRs have already found several applications, including lasing~\cite{Hakala2018} and second-harmonic generation~\cite{Michaeli2017b,Hooper2018}.

Although SLR-supporting metasurfaces are easily designed and fabricated, their post-fabrication control remains challenging. 
Earlier, control has been achieved, for example, by straining the elastic substrate of the metasurface,\cite{Gupta2019} using refractory materials and raising the ambient temperature to very high values\cite{Zakomirnyi2017}, or by using optically~\cite{Taskinen2020} or thermally~\cite{Volk2017} responsive polymers. Furthermore, temperature-dependent optical responses of plasmonic waveguides~\cite{gerasimov2017} and LSPRs~\cite{gerasimov2017, Yeshchenko2013b, Bouillard2012} have been already studied. However, control over high-$Q$ SLRs ($Q\geq500$) have not yet been investigated.

Here, we demonstrate dramatic control over the $Q$-factor and the extinction of a high-$Q$ SLR by breaking the symmetry of the surroundings via temperature-dependent substrates. Unlike previous work, our devices operate close to room temperature making our approach broadly applicable. 
We demonstrate how a decrease of \SI{10}{\celsius} in the ambient temperature of the devices result in an increase of the $Q$-factor from 400 to 750, a 15\% increase in the extinction and a 1\,nm shift in the center wavelength $\lambda_c$ of the SLR. 
Our results demonstrate accurate and reversible modification of the spectral properties of SLRs, paving the way towards tunable SLR-based photonic devices. 
More generally, our results demonstrate how highly efficient and potentially ultrafast tuning of SLR properties could be achieved by breaking the symmetry of the dielectric environment.

The properties of LSPRs are central in the field of plasmonics. The properties depend on the size, shape, environment and the material of the NP.\cite{Maier2007}
The LSPRs dictate the optical properties of plasmonic NPs, which can be understood by introducing the polarizability of a NP $\alpha$ and by connecting it to the induced dipole moment $\vec{p}$ and to the incoming field $\vec{E}$ via\cite{Maier2007} 
\begin{equation}
\vec{p} = \varepsilon_0 \varepsilon_s \alpha \vec{E} \,, \label{eq:dipolemoment}
\end{equation} 
where $\varepsilon_0$ and $\varepsilon_s$ are the vacuum and surrounding medium permittivities, respectively. The details of the calculation of the polarizability $\alpha$ are given in the Supporting Information.  

LSPRs suffer from broad linewidths and low $Q$-factors induced by the high ohmic losses typical for metals. The $Q$-factors of plasmonic metasurfaces have been successfully improved by arranging the NPs in periodic lattices and utilizing the emerging diffractive properties~\cite{Auguie2008}. In such structures, the diffraction modes can couple with the individual LSP modes giving rise to collective responses known as SLRs. 

The formation of SLRs can be calculated with a semi-analytical method known as lattice-sum approach (LSA).\cite{Huttunen2016a} In the LSA, the effects of scattered fields affecting the NPs are reduced to a single variable, called the lattice sum $S$. In homogeneous environment, the lattice sum can be derived using the free-space dyadic Green's tensor (see the Supporting Information for more information). 

A NP near a planar interface between materials with different refractive indices show modified dipole radiation pattern with a large portion of the energy emitted into the optically denser material \cite{lieb2004}. This hinders the formation of the diffraction mode, and ultimately, the intensity and line width of the SLR \cite{Khlopin2017}. Under these circumstances, we can also expect the diffractive modes to form independently in each material, in which case the heterogeneous lattice sum can be estimated to be
\begin{equation}\label{eq:latticesum2}
S_{\mathrm{het}}(n_1(T),n_2(T)) = \frac{S(n_1(T))+S(n_2(T))}{2}, 
\end{equation}
where $S(n)$ is the lattice sum in homogeneous environment, while $n_1$ and $n_2$ are the refractive indices of the substrate and superstrate, respectively. 
The LSA uses the lattice sum $S$ to modify polarizability $\alpha$ to include the effects of the scattered fields on a single NP. Once taken into account, the effective polarizability $\alpha^*$ of a NP can be written as
\begin{equation}
    \alpha^*(T) = \frac{1}{1/\alpha - S(T)} \, .
\end{equation}
Note that the inspection here is only valid for one polarization type. A tensorial approach is needed for more general results~\cite{Huttunen2016a}. Once the effective polarizability $\alpha^*$ has been solved, the transmission through the metasurface can be calculated using\cite{Reshef2019}:
\begin{equation}
    \sigma_{\text{trans}}  = 1-\frac{ k}{P_xP_y}\text{Im}[\alpha^*],
\end{equation} 
where $P_x$ and $P_y$ are the periodicities of the NP array in respective dimensions.

When the diffractive modes form separately in the different materials, as estimated by Eq.~\eqref{eq:latticesum2}, already small differences in the refractive indices of the materials can cause drastic changes in the $Q$-factors and extinction spectra of the measured devices. This occurs because the diffractive mode -conditions shift to different wavelengths for the substrate and superstrate, consequently broadening the line width of the measured extinction peak. 

The difference between the refractive indices of the substrate and superstrate is controlled in this work via temperature. In addition to affecting the environment, temperature has a small but undeniable effect also on the responses of individual NPs, which was investigated numerically in the Supporting Information. Although the small temperature changes used in this work were estimated to result in negligible changes in the optical responses of individual NPs, we note that it would be interesting to study whether cryogenic temperatures could facilitate realizations of ultra-high-$Q$ SLRs.
 
\begin{figure}
    \centering
    \includegraphics{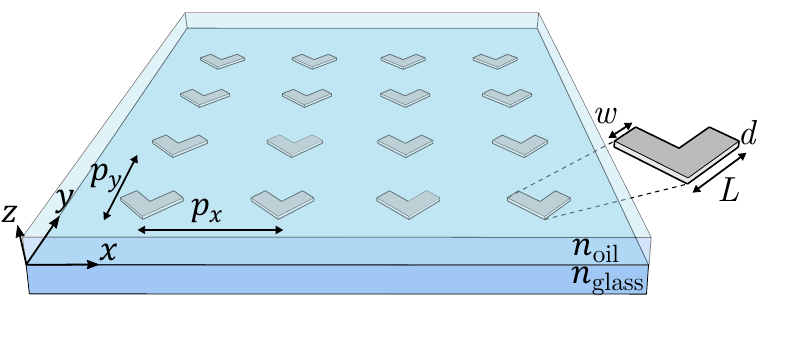}
    \caption{The investigated metasurface consisting of V-shaped aluminum NPs fabricated on a glass substrate ($n_{\mathrm{glass}}=1.511$) and covered by immersion oil
    . Here,  $w= 70$ nm, $L= 140$ nm, and $d= 30$ nm are the arm widths, arm lengths, and the thicknesses of the NPs. The NPs were arranged in a rectangular lattice ($p_x=727$ nm and $p_y=400$ nm), giving rise to surface lattice resonances near 1100 nm for $y$-polarized light.}
    \label{fig:Sample}
\end{figure}

In this work, we investigated an array of V-shaped aluminum NPs with periodicities of \SI{727}{nm} in $y$-direction and \SI{400}{nm} in $x$-direction (see Fig.~\ref{fig:Sample}). The NPs were \SI{30}{nm} thick with \SI{140}{nm} long and \SI{70}{nm} wide arms. The NPs were fabricated using electron-beam lithography on a 1\,mm thick D263T glass substrate from Schott. Olympus type-F immersion oil was used to surround the particles in a matching refractive index environment, and a cover glass with an anti-reflective coating for 1000--1300\,nm wavelengths was placed on top to complete the sample. The anti-reflective coating was used to avoid Fabry--Pérot resonances from multiple reflections from different interfaces. For more information, see the Supporting Information.

\begin{figure*}
    \centering
    \includegraphics{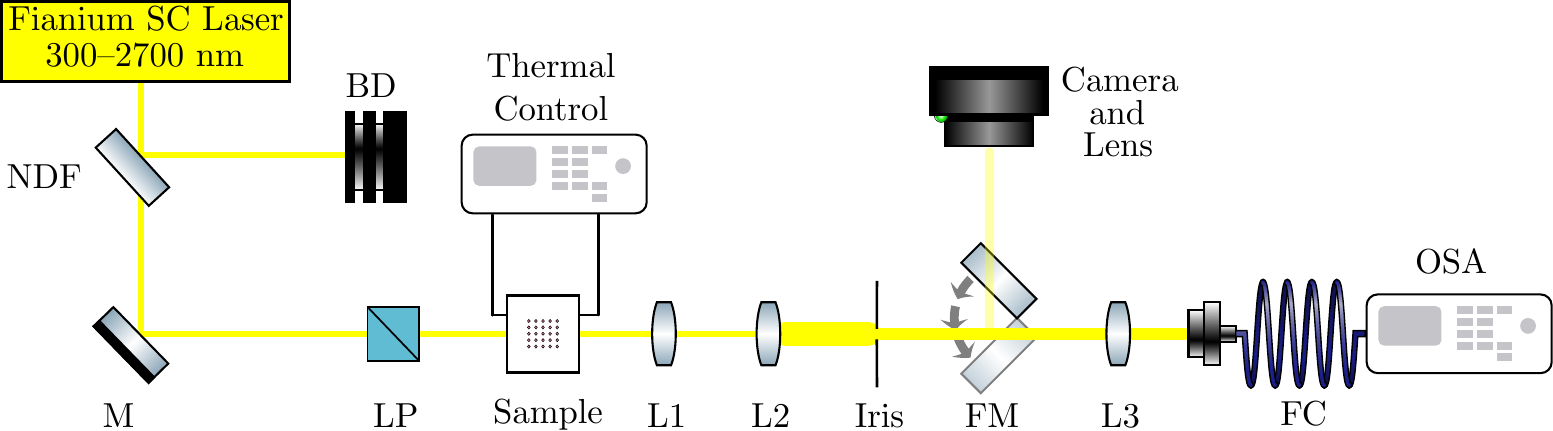}
    \caption{Schematic representation of the experimental setup used to measure the transmission spectra and to control the ambient temperature of the studied structures. SC, supercontinuum; NDF, neutral-density filter; BD, beam dump; M, mirror; LP, linear polarizer; L, lenses (L1, $f$ = 19 mm; L2, $f$ = 75 mm; L3, $f$ = 4.6 mm); FM, flip mirror; FC, fiber coupler; OSA, optical spectrum analyzer; SM, spectrometer.}
    \label{fig:setup}
\end{figure*}
The experimental setup is shown in Figure \ref{fig:setup}. A Fianium supercontinuum laser was used as a broadband light source with wavelength range of 300--2700 nm and maximum power $<$50 W.

The supercontinuum power was kept at a suitable level ($<$\SI{20}{mW}) using a neutral-density-filter in order to keep the possible absorption-based heating of the sample negligible. The ambient temperature $T$ of the sample was controlled with a standard thermoelectric cooler, connected to an adjustable voltage source for fine-tuning of the temperature. The temperature $T$ of the sample was determined using a Flir E85 thermal imaging camera. A linear polarizer before the sample ensured the sample excitation using linearly polarized light. An iris after the sample was used to limit the collection of light to include only light passed through the desired metasurface array. Finally, the beam of light was guided to an optical spectrum analyzer through a single-mode optical fiber. 

\begin{figure*}[h]
    \centering
    \begin{tikzpicture}

%%%% NODES
 \node[] at (6.9,2.9) {\textbf{d)}};
 \node[] at (6.9,5.7) {\textbf{c)}};
 \node[] at (6.9,9.2) {\textbf{b)}};
 \node[] at (-1.5,9.2) {\textbf{a)}};

\xdef\figXscaler{0.3}
 \xdef\figYscaler{0.4}

\begin{scope}[shift={(-0.3,0)}]

\begin{axis}[                       %axis 1
        /pgf/number format/.cd,
       % use comma,
        1000 sep={},
    restrict x to domain=1075:1125,
    axis lines = box,
    x=\figXscaler*0.9cm/2,
     y=\figYscaler*13cm/1, 
    axis line style = {thick},
    every axis plot/.append style={thick},
    xlabel = {$\lambda$ (nm)},
    ylabel = {Transmission (\%)},
    xmin = 1075, xmax = 1125, ymin = 0.4, ymax = 2.225,
    xtick={1080,1090,1100, 1110, 1120},
    ytick={0.5,0.6,...,2.2},
    %ytick={}
    %y tick label style={/pgf/number format/.cd, fixed,fixed zerofill,precision=1,/tikz/.cd},
    yticklabels={50,60,...,100,70,80,90,100,70,80,90,100,70,80,90,100},
    legend pos=south west,
    grid style={dotted,gray},
    grid=both,
    %ymajorgrids=true,
    %grid style=dashed,
    % width = 8cm,
]

    \addplot[mark=none, gray, thick,forget plot] coordinates {(1075,1) (1125,1)};
    \addplot[mark=none, gray, thick,forget plot] coordinates {(1075,1.4) (1125,1.4)};
    \addplot[mark=none, gray, thick,forget plot] coordinates {(1075,1.8) (1125,1.8)};
    \addplot[mark=none, gray, thick,forget plot] coordinates {(1075,2.2) (1125,2.2)};
    
    \addplot[
    color = blue,
    mark size=0.5pt,
    ]
table[]
{Data/11K_D4_verticallyrised.dat};
\addlegendentry{$11^{\circ}$C}
\addplot[
    color = cyan,
    mark size=0.5pt]
table[]
{Data/21K_D4_verticallyrised.dat};
\addlegendentry{$21^{\circ}$C}
\addplot[
    color = pink,
    mark size=0.5pt]
table[]
{Data/31K_D4_verticallyrised.dat};
\addlegendentry{$31^{\circ}$C}
\addplot[
    color = red,
    mark size=0.5pt]
table[]
{Data/39K_D4_verticallyrised.dat};
\addlegendentry{$39^{\circ}$C}

    \addplot[
    color = black,
    mark size=0.5pt,
    ]
table[]
{Data/284K_LSA_verticallyrised.dat};
    \addplot[
    color = black,
    mark size=0.5pt,
    ]
table[]
{Data/294K_LSA_verticallyrised.dat};
    \addplot[
    color = black,
    mark size=0.5pt,
    ]
table[]
{Data/304K_LSA_verticallyrised.dat};
    \addplot[
    color = black,
    mark size=0.5pt,
    ]
table[]
{Data/312K_LSA_verticallyrised.dat};

\end{axis}
\end{scope}

%%%%%%%%%%%%%%%%%%%%%%%%%%%%
    
\begin{scope}[shift={(8.6,0)}]
\begin{axis}[                   %AXIS3
        /pgf/number format/.cd,
        use comma,
        1000 sep={},
    axis lines = box,
    x=\figXscaler*0.9cm/2,
     y=\figYscaler*1cm/130,
    axis line style = thick,
    %every axis plot/.append style={thin},
    xlabel = {$T$ (\SI{}{\celsius})},
    ylabel = {$Q$},
    xmin = 6, xmax = 41 , ymin = 100, ymax = 1150,
    legend pos=north west,
    xtick = {10,15,...,40},
    ytick = {150, 300,...,1050},
    legend pos=north east,
]

    \addplot[
    only marks,
    mark = x,
    color = red,
    mark size=3pt,
    style=ultra thick]
table[]
{Data/exp_Q.dat};
\addplot[
    %only marks,
    %mark = o,
    color = blue,
    mark size=3pt,
    style= thick]
table[]
{Data/LSA_Q.dat};

\end{axis}
\end{scope}

\begin{scope}[shift={(8.6,\figYscaler*8.5)}]

\begin{axis}[                           %AXIS3
        /pgf/number format/.cd,
        use comma,
        1000 sep={},
        %restrict x to domain=9.5:40.5,
    axis lines = box,
    x=\figXscaler*0.9cm/2,
     y=\figYscaler*1cm/1,
    axis line style = thick,
    every axis plot/.append style={ thick},
    ylabel = {$\lambda_c$ (nm)},
    xticklabel =\empty,
    xmin = 6,xmax = 41 , ymin = 1095.5, ymax = 1101.5,
    ytick={1096,1097,...,1101},
    xtick={10,15,...,40},
    legend pos=north west,
    legend pos=north east,
]
 %   \addplot[samples=50, domain=8:42,color = gray,style = dashed, style=thin] {-0.1125 *x+1101.1 };
    \addplot[
    only marks,
    mark = x,
    color = red,
    mark size=3pt,
    style = ultra thick]
table[]
{Data/exp_peak.dat};

    \addplot[
    %only marks,
    %mark = o,
    color = blue,
    mark size=3pt]
table[]
{Data/LSA_peak.dat};
\end{axis}
\end{scope}

\begin{scope}[shift={(8.6,\figYscaler*15)}]

\begin{axis}[                           %AXIS3
        /pgf/number format/.cd,
        use comma,
        1000 sep={},
    axis lines = box,
    x=\figXscaler*0.9cm/2,
     y=\figYscaler*2cm/10,
    axis line style = thick,
    every axis plot/.append style={thick},
    ylabel = {Extinction (\%)},
    xticklabel =\empty,
    xmin = 6,xmax = 41, ymin = 6, ymax = 50,
    ytick={10,15,...,45},
    xtick={10,15,...,40},
    legend pos=north west,
    legend pos=north east,
]
    \addplot[
    only marks,
    mark = x,
    color = red,
    mark size=3pt,
    style = ultra thick]
table[]
{Data/exp_intensity.dat};

    \addplot[
    %only marks,
    %mark = o,
    color = blue,
    mark size=1pt]
table[]
{Data/LSA_intensity.dat};
\end{axis}
\end{scope}
\end{tikzpicture}
    \caption{a) The transmission spectra measured at different ambient temperatures $T$. The black lines represent respective LSA calculations. The SLR peak at \SI{39}{\celsius} with central wavelength of \SI{1097}{nm} and a line width of \SI{5}{nm} gets narrower and redshifted with decreasing $T$ down to a line width of \SI{1}{nm} and central location of \SI{1100}{nm} at \SI{11}{\celsius}. The b) peak extinction, c) spectral location $\lambda_c$, and d) $Q$-factor of the resonance as a function of temperature $T$. The experimental data is marked with red crosses, and the LSA calculations with solid blue lines. }
    \label{fig:experimental_data}
\end{figure*}
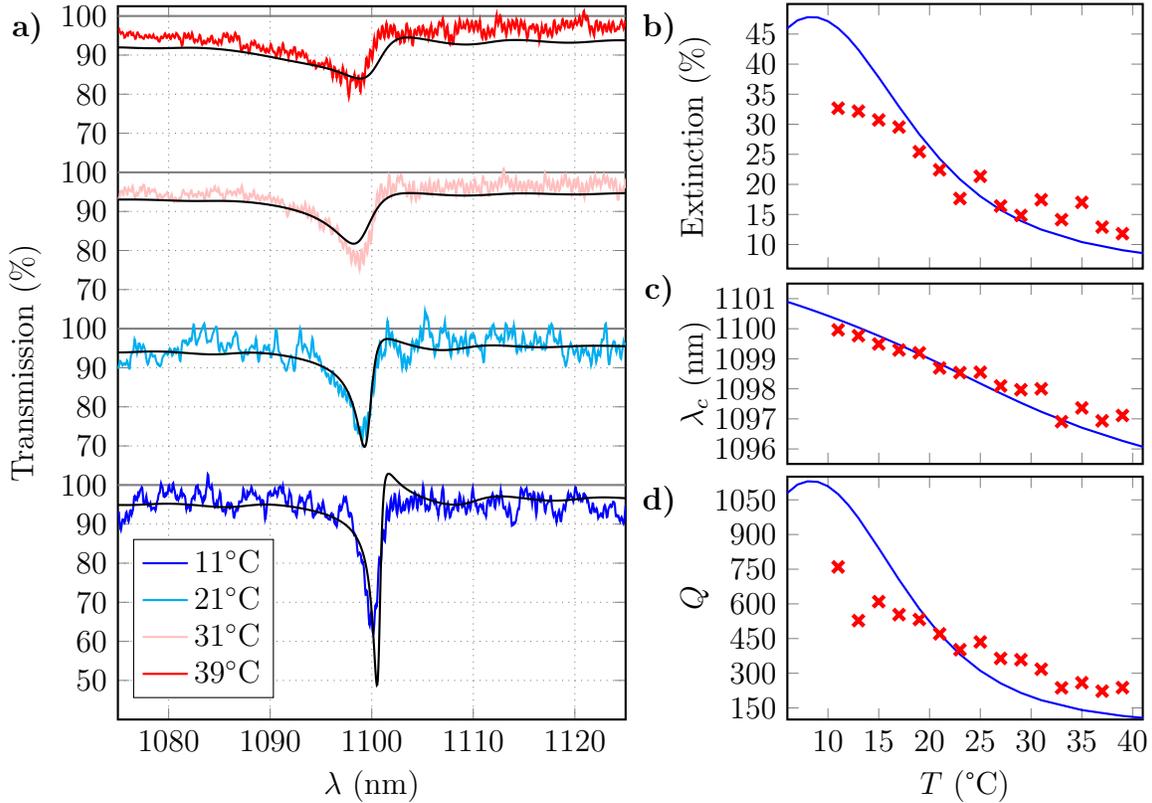

The measured transmission spectra with ambient temperature ranging from \SI{11}{\celsius} to \SI{39}{\celsius} are shown in Fig.~\ref{fig:experimental_data}a. The results are in good agreement with those extracted from semi-analytical LSA calculations (black lines). We note that the LSA calculations are readily extendable to higher temperature ranges (see the Supporting Information for more details on LSA calculations). 

The $Q$-factors, peak extinctions, and the center wavelengths $\lambda_c$ of the peak were obtained by fitting a Lorentzian line shape to the experimental data (see the Supporting Information for fitted models). The resonance showed a decreasing $Q$-factor and extinction, and a minor blueshift with increasing temperature. A linear regression for spectral location data showed an average decrease in $\lambda_c$ of \SI{0.11}{nm/\celsius}. This result can be attributed mainly to the decrease in the refractive index of the immersion oil with increasing $T$. Strikingly, the $Q$-factor and extinction both showed a 3.5-fold decrease in magnitude in the temperature range from \SI{11}{\celsius} to \SI{39}{\celsius}. 
The strong responsiveness to temperature is a result of symmetry breaking of the refractive index profile of the dielectric environment. The shift in peak location $\lambda_c$ has also a minor effect on both the $Q$-factor and extinction as the energetic overlap between plasmonic and diffractive modes increases with rising temperature.

The behaviour of the peak extinction, location, and $Q$-factor are shown in Fig.~\ref{fig:experimental_data}b--d, with the experimental data marked with red. The extended temperature range of the LSA calculations in blue show that both the $Q$-factor and the extinction should reach their maximum values at around \SI{8}{\celsius}. At this temperature, the NP surroundings become symmetric ($n_{\rm{oil}}=n_{\rm{glass}}$). With increasing $T$, the refractive index of the index matching oil decreases faster ($\dv*{n_{\rm{oil}}}{T}=\,$\SI{-3.3d-4}{1/\celsius}) than that of the glass substrate ($\dv*{n_{\rm{glass}}}{T}\approx\,$\SI{-6d-6}{1/\celsius}), resulting in symmetry breaking of the environment. We expect the difference in the refractive indices ($n_{\rm{glass}}-n_{\rm{oil}}$) to range from \SI{0}{} at around \SI{8}{\celsius} to around \SI{1d-2}{} at \SI{40}{\celsius} in an almost linear manner. The effects of symmetry breaking in the $Q$-factor and extinction are therefore confirmed by the LSA calculations (see Figs.~\ref{fig:experimental_data}b,d and the Supporting Information).  

While the $Q$-factor and extinction of the SLR are expected to rise even more with lower temperatures, the humidity of our lab limited our measurements to \SI{11}{\celsius}. At colder temperatures water condensation to the sample surfaces made further experiments unfeasible.
Nevertheless, our results still show dramatic changes in the SLR properties already with quite small changes in the ambient temperature. Furthermore, we operate the devices close to room temperatures, which demonstrates that the approach could be an easily and broadly applicable tuning method. 

Our results confirm the significance of symmetric environment in high-$Q$ SLR metasurfaces. We estimate that already a modest change of $\sim$\SI{0.003}{} units in the refractive indices of the super- and substrate (from temperatures \SI{21}{\celsius} to \SI{11}{\celsius}) raised the $Q$-factor of the resonance from 400 to 750. Furthermore, the results show that the spectral properties of SLRs can be efficiently controlled via symmetry breaking in the sample, which is particularly interesting when noting that alternative means of changing the refractive index of the superstrate could be used in a manner similar to the temperature control. An ultra-fast alternative could be to utilize Kerr-active materials, making it possible to control the properties of the SLRs by using an external voltage source or by using a control beam of light. 

To conclude, we have investigated how spectral properties of plasmonic surface lattice resonances can be modified by controlling the ambient temperature of the studied metasurface. Our metasurface consisted of aluminum nanoparticles arranged in a rectangular array on a glass substrate covered by immersion oil. At room temperature, the metasurface exhibit a high quality factor ($Q\approx400$) SLR near \SI{1100}{nm}. By decreasing the ambient temperature by \SI{10}{\celsius}, the SLR peak was slightly redshifted and the $Q$-factor was increased to 750. The increased $Q$-factor is explained by the improved symmetry of the nanoparticle surroundings, resulting from the temperature-dependent refractive index of the overlying immersion oil. Our results show that even slight changes in the refractive indices of the surrounding materials can result in dramatic changes in the SLR properties, simultaneously demonstrating their accurate and reversible tunability.

\section{Supporting Information}
The Supporting Information is provided at the end of this submission.

\section{Acknowledgements}
The authors acknowledge the support of the Academy of Finland (Grant No. 308596), the Flagship of Photonics Research and Innovation (PREIN) funded by the Academy of Finland (Grants No. 320165 and 320166). 
TS acknowledges also Jenny and Arttu Wihuri Foundation for doctoral research grant. TKH acknowledges Academy of Finland project number (322002).
The authors thank Jarno Reuna for providing anti-reflection coatings.

\nocite{Barnes2016}
\nocite{Jensen1999}
\nocite{Zoric2011}
\nocite{Rakic1995}
\nocite{Huttunen2016a}
\nocite{SCHOTT}
\nocite{SCHOTTtemperature}
\nocite{olympus}
\nocite{Unser2015}
\nocite{Bouillard2012}
\nocite{Johnson1972}
\nocite{CRC}
\nocite{Bouillard2012}
\nocite{kittel2004}
\bibliography{NLOmetasurfaces.bib}

\onecolumn

{\Large \centering \textbf{Supporting Information}}

\section{Sample Fabrication}
\label{SM:fabrication}
For this work, we fabricated V-shaped aluminum (Al) nanoparticle arrays with a total area of $300\times300$ \si{\micro\m^2}.
 The structures were fabricated on a pre-cleaned microscope slide (Schott Nexterion, D263T glass). A 200~nm layer of PMMA-resist (MicroChem, 950k) was spin-coated on top and baked on a hot plate at 180$^\circ$C for 180~s. A 10~nm layer of Al was evaporated on the resist to act as a conductive layer for electron beam lithography.
 
The patterning was done using a Raith EBPG 5000+ 100~kV electron beam lithography system. After patterning the Al layer was removed using a 1\% sodium hydroxide solution. The resist was then developed using a 1:3 mixture of methyl isobutyl ketone and isopropanol (IPA) for 15~s, followed by a 30~s immersion in IPA. The sample was dried with nitrogen and placed in an electron beam evaporator for depositing 30~nm of Al. Finally, a liftoff process was performed by soaking the sample in acetone overnight and gently washing the surface with more acetone using a syringe. This removes the resist and excess metal on top of it, leaving only the nanoparticles on the glass substrate. The sample was then rinsed with IPA and dried with nitrogen.

Before the measurements, we covered the metasurface with index-matching oil and an anti-reflection (AR) coated coverslip with the AR wavelength band at 1000--1300 nm. This way, the nanoparticles were assured to have a homogeneous surrounding, and we avoided any Fabry--P\'erot resonances resulting from multiple reflections from different interfaces present in the fabricated devices. 

\section{Localized Surface Plasmon Resonance}
\label{SM:LSPR}
Polarizability for NPs of arbitraty shape is given by~\cite{Barnes2016}
 \begin{equation}\label{eq:polarizability}
    \alpha_i = V\frac{\varepsilon_m - \varepsilon_s}{\varepsilon_s + L_i(\varepsilon_m-\varepsilon_s)} ,
\end{equation}
where $V$ is the volume of the nanoparticle (NP), $\varepsilon_m$ and $\varepsilon_s$ are the permittivities of the metal and surrounding media respectively, $L_i$ is a factor depending on the geometry of the particle, and index $i$ marks the relevant dimension. While $L_i$ has been analytically solved for shapes such as slabs and ellipsoids~\cite{Barnes2016}, a general analytical solution for complicated shapes, such as the V-shaped particles used in this work, does not exist. Consequently, we used a value of $L_y = 0.26$, that was found by trial and error to match well with the experiments. 

Equation \eqref{eq:polarizability} assumes the electric field to be static at a given time, which produces accurate results while the dimensions of the NPs are smaller than 1\% of the incident wavelength \cite{Jensen1999}. With our NPs, however, a modified long-wavelength approximation (MLWA) using dynamic perturbations must be used for accurate modelling of the localized surface plasmon resonances (LSPRs). After applying MLWA, polarizability 
can be written as \cite{Zoric2011}
\begin{equation} \label{eq:MLWA}
\alpha_{i, \mathrm{MLWA}}=\frac{\alpha_{i}}{1-\mathrm{i}\frac{k^3}{6\pi}   \alpha_{i}-\frac{k^{2}}{4\pi a_{i}} \alpha_{i}},
\end{equation}
where $\alpha_i$ is the static polarizability, and $k$ is the wavenumber of the incident field. 

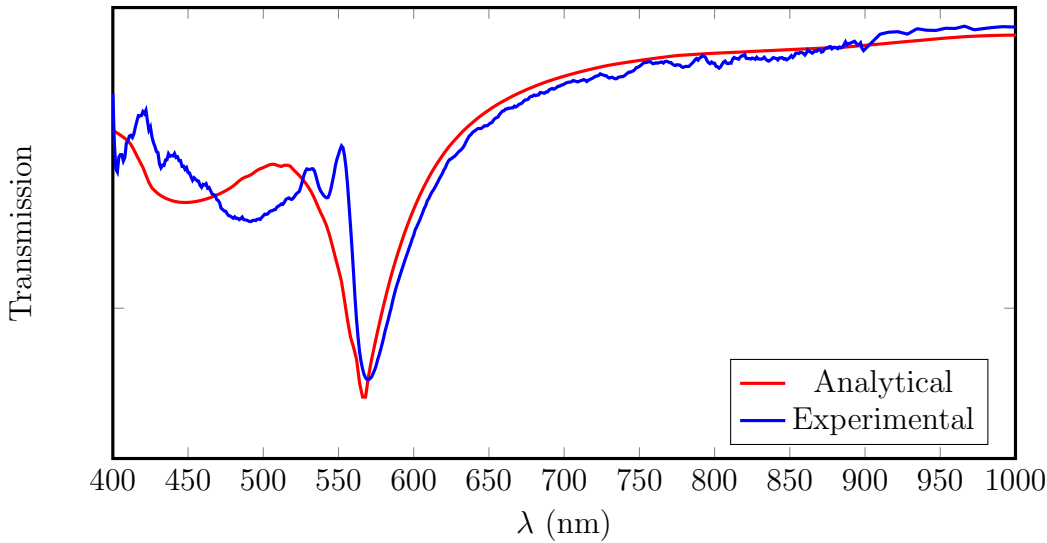
\begin{figure}
    \centering
    \begin{tikzpicture}
%%%% NODES
\xdef\figXscaler{0.3}
 \xdef\figYscaler{0.4}

\begin{scope}[shift={(0,0)}]

\begin{axis}[                       %axis 1
/pgf/number format/.cd,
        use comma,
        1000 sep={},
    axis lines = box,
    x=\figXscaler*1cm/15,
     y=\figYscaler*50cm/1, 
    axis line style = very thick,
    every axis plot/.append style={very thick},
    xlabel = {$\lambda$ (nm)},
    ylabel = {Transmission},
    xmin = 400, xmax = 1000, ymin = 0.7, ymax = 1,
    ytick={0.6,0.8,...,2.4},
    legend pos=south east,
    yticklabels={,,},
    %ymajorgrids=true,
    %grid style=dashed,
    % width = 8cm,
]

    \addplot[
    color = red]
table[]
{Data/LSPR_LSA.dat};
\addlegendentry{Analytical}
    \addplot[
    color = blue]
table[]
{Data/LSPR_mes.dat};
\addlegendentry{Experimental}
\end{axis}
\end{scope}

\end{tikzpicture}
    \caption{Measured LSPR of the metasurface used in the work (blue). The LSPR given by the LSA calculations used in this work (red).}
    \label{fig:LSPR_LSA}
\end{figure}

Figure \ref{fig:LSPR_LSA} shows experimental data next to analytical model using equations \eqref{eq:polarizability} and \eqref{eq:MLWA}. The experimental data is measured from the same metasurface as the SLRs in this work. It should be noted that the LSPR is not pure due to the presence of the second order diffraction-mode near \SI{550}{nm}.  In the equation \eqref{eq:polarizability}, experimental data for aluminum permittivity from Rakić \cite{Rakic1995} was used for $\varepsilon_m$. The surroundings were modelled using the permittivity of the SCHOTT - multiple purpose D 263® T eco Thin glass. In the Equation \eqref{eq:MLWA}, a NP dimension of $a_y = \SI{75}{nm}$ was used, corresponding approximately to half of the length of the NP in $y$-direction. In Equation \eqref{eq:polarizability} a geometrical factor a of $L_y = 0.26$ was used.

\section{Lattice Sum Approach}
Lattice sum approach (LSA) method was used to support the experimental results with semianalytical calculations. LSA is a simplified version of the discrete-dipole approximation (DDA) method that allows to calculate the response of optically coupled nanoparticles. In LSA, the effects of coupling between nanoparticles are reduced to a single constant for all nanoparticles. The reduction is done with the loss of generality in non-infinite lattices, with LSA being unable to model dipole moments of the nanoparticles near the edges of the lattice. The LSA is based on several assumptions. Most importantly, all the nanoparticles are treated as point-like objects which scatter the incoming light as electric dipoles. In LSA specifically, it is assumed that all the nanoparticles are identical, and experience identical environments with respect to each other.

In LSA, a concept of effective polarizability $\alpha^*$ is introduced, which associates the effectively induced dipole moments $\vec{p}$ to the incident electric field $\vec{E}$ via $\vec{p} = \varepsilon_0\varepsilon_s \alpha^* \vec{E}$, where $\varepsilon_0$ and $\varepsilon_s$ are the vacuum permittivity and relative permittivity of the surroundings, respectively. The effective polarizability is given by
\begin{equation}\label{eq:effective_polarizability}
    \alpha^*_i = \frac{1}{1/\alpha_i - S_i},
\end{equation}
where $i$ denotes a component of the quantity in Cartesian direction, $\alpha_i$ is the single particle polarizability, and $S_i$ is the lattice sum.

The lattice sum in equation \eqref{eq:effective_polarizability} can be expressed in a homogeneous environment as \cite{Huttunen2016a}
\begin{equation}\label{eq:latticesum}
S_{i}(k)=\sum_{j=1}^{N} \frac{\rme^{\rmi k r_{j}} }{4 \pi r_{j}}\left[k^{2} \sin ^{2} \varphi_{i,j}+\frac{\left(1-\rmi k r_{j}\right)\left(3 \cos ^{2} \varphi_{i,j}-1\right)}{r_{j}^{2}}\right],
\end{equation}
where $N$ is the number of particles taken into account, $r_j$ is the distance between $j^{\mathrm{th}}$ particle and the center particle, and $\varphi_{i,j}$ is the angle between the $i^{\mathrm{th}}$ dipole moment component and the vector from the center particle to $j^{\mathrm{th}}$ particle. The lattice sum $S$ can be derived from DDA by assuming identical dipole moments for all NPs~\cite{Huttunen2016a}.

In this work, the studied NP arrays lied on an interface between Olympos IMMOIL-F30CC -immersion oil and SCHOTT - multiple purpose D 263® T eco Thin Glass. This caused slight heterogeneity in the surroundings of the SLR at room temperature. To model the heterogeneity, the coupling of NPs was thought to develop independently in the different materials.
We assumed the heterogeneous lattice sum to be an arithmetic mean of two different homogeneous lattice sums arising in the different surrounding materials. 
Particularly, the total lattice sum was calculated as follows
\begin{equation}\label{eq:latticesum2}
S_{i,\mathrm{het}}(k_{\mathrm{oil}},k_{\mathrm{glass}}) = \frac{S_i(k_{\mathrm{oil}})+S_i(k_{\mathrm{glass}})}{2}, 
\end{equation}
where $k_{\mathrm{oil}}$ and $k_{\mathrm{glass}}$ are the wavenumbers in the (Olympos IMMOIL-F30CC -immersion) oil and (SCHOTT - multiple purpose D 263® T eco Thin) Glass respectively, given by
\begin{equation}
    k_{\mathrm{oil}} = \frac{2\pi n_{\mathrm{oil}}(\lambda,T)}{\lambda}, \qquad k_{\mathrm{glass}} = \frac{2\pi n_{\mathrm{glass}}(\lambda,T)}{\lambda}
\end{equation}
where $\lambda$ is the wavelength, and $n_{\mathrm{oil}}(\lambda,T)$ and $n_{\mathrm{glass}}(\lambda,T)$ are the refractive indices of the oil and glass respectively. Dispersion data for $n_{\mathrm{glass}}(\lambda,T)$ was taken from \cite{SCHOTT}. With the lack of full description of temperature dependence in SCHOTT - multiple purpose D 263® T eco Thin Glass, a temperature dependence of a glass with similar properties (SCHOTT, N-KF9) was used \cite{SCHOTTtemperature}. $n_{\mathrm{oil}}(\lambda,T)$ was calculated from the constants $\rmd n/\rmd T = -3.3\times10^{-4}$ and Abbe number $V_e = 41$ given by the manufacturer~\cite{olympus}. With the immersion oil not designed to operate at infrared wavelengths,  we added a free parameter to match the calculations better with experiments in longer wavelengths. Namely a constant of $n_0 = 0.035$ was added to the refractive index as follows: $n_{\mathrm{oil,used}}(\lambda,T) = n_{\mathrm{oil}}(\lambda,T)+n_0$. 

It should be noted that while in both surrounding materials the refractive index decreases with increasing temperature, the effect is roughly one magnitude stronger in the oil superstrate compared to the glass substrate. This discrepancy causes the heterogeneity in the environment of the NPs to rapidly rise with increasing temperature. In the calculations this shows up first as widened peaks in both the real and imaginary parts of the lattice sum $S_{i,\mathrm{het}}$ near the diffraction orders associated with the lattice. At high enough temperatures, two separate peaks will form, proving the method demonstrated by the equation \eqref{eq:latticesum2} insufficient for modelling high heterogeneity in the surroundings of the NPs. However, good correlation between the performed LSA calculations and the experimental data was observed in the narrow temperature range studied in our work.

While the heterogeneity of the surroundings is not the only factor affecting the SLR, it is the most important one regarding $Q$-factor and extinction. Other factors affecting the SLR-properties include the energetic overlap between the plasmonic and diffractive modes, with higher $Q$-factors effectively achieved with smaller values of polarizability $\alpha$ at the diffraction mode wavelength. Ways to modify the polarizability $\alpha$ include control over the size, shape and material of the NP, and control over the refractive index of the surrounding medium (see Equation \eqref{eq:polarizability}). Post-fabrication control over the single NP properties therefore usually depend on changing the refractive index of the surrounding medium (such as in biosensing with plasmonic NPs~\cite{Unser2015}). Additionally, temperature has an effect on the permittivity of the metal, and therefore on the single NP properties. These effects have been investigated in the following section.

\section{Effects of temperature on the permittivity of the metal}
\label{SM:NP_Tdependence}
It is difficult to model the temperature dependence of the aluminum permittivity using the Drude-model near visible and infra-red frequencies due to interband excitations near 800 nm. 
However, since the experiments were carried out near the wavelength of 1100 nm, the interband excitations taking place near 800 nm can be expected to play a negligible role. Nevertheless, we decided to estimate the temperature dependence of the permittivity of metallic NPs, and subsequent changes in the SLRs of periodic nanoparticle arrays, for gold NPs. This approach was chosen because the permittivity of aluminum can be expected to behave very similarly to that of gold near the wavelength of 1100 nm.  
The permittivity of gold was modelled using the Drude model \cite{Bouillard2012}:
\begin{equation}
\varepsilon_m(T)=\varepsilon_{\infty}-\frac{\omega_{p}^{2}(T)}{\omega(\omega+i \gamma(T))},
\end{equation}
where $\varepsilon_m$ is the permittivity of gold, $\varepsilon_{\infty}$ is the high-frequency permittivity, $\omega_p$ is the plasma frequency of the metal, $\omega$ is the angular frequency of the incident light, and $\gamma$ is the relaxation constant for the free electrons. Plasma frequency is given by \cite{Bouillard2012}
\begin{equation}
    \omega_p(T) = \sqrt{\frac{n_oe^2}{\varepsilon_0m(1+3\beta\Delta T)}},
\end{equation}
where $n_0$ is the free electron number density in the metal, $e$ is the elementary charge, $\varepsilon_0$ is the vacuum permittivity, $m$ is the mass of electron, $\beta$ is the linear thermal expansion coefficient and $\Delta T = T-T_0$ is the difference to the temperature of comparison $T_0$. The term $1+3\beta \Delta T$ arises from the thermal expansion of the material, which is the temperature dependent effect affecting the plasma frequency.

The thermal relaxation constant $\gamma$ is comprised of electron--electron contribution $\gamma_{\mathrm{e-e}}(T)$ and electron--phonon contribution $\gamma_{\mathrm{e-ph}}(T)$, which are given by \cite{Bouillard2012}
\begin{equation}
\begin{aligned}
\gamma_{e-e}(T) &=A\left[\left(k_{B} T\right)^{2}+(\hbar \omega)^{2}\right] ,\\
\gamma_{e-ph}(T) &=\gamma_{0}\left[\frac{2}{5}+\frac{4 T^{5}}{\theta_{D}^{5}} \int_{0}^{\theta_{D} / T} \frac{x^{4}}{e^{x}-1} \mathrm{~d} x\right],
\end{aligned}
\end{equation}
where the factor $A$ is dependent on the properties of the conduction band of the metal \cite{Bouillard2012}, $k_B$ is the Boltzmann constant, $\hbar$ is the reduced Planck's constant, $\gamma_0$ is the temperature-independent electron--phonon relaxation constant and $\Theta_D$ is the Debye-temperature. $\varepsilon_\infty=$ \SI{11.5}{} and $\hbar \gamma_0=$ \SI{0.07}{eV} were determined by fitting gold bulk permittivity (from Johnson \& Christy \cite{Johnson1972}) to the model. Other constants for gold are $n_0 =$ \SI{5.9d28}{m^{-3}}, $\beta =$ \SI{14.2d-6}{K^{-1}}~\cite{CRC}, $\hbar A=$ \SI{0.0317}{eV^{-1}}~\cite{Bouillard2012} and $\Theta_D = $ \SI{170}{K} ~\cite{kittel2004}.

\begin{figure}
    \centering
    \begin{tikzpicture}
%%%% NODES
\node[] at (7.1,2.6) {\textbf{c)}};
\node[] at (7.1,6) {\textbf{b)}};
\node[] at (-1,6) {\textbf{a)}};
\xdef\figXscaler{0.3}
 \xdef\figYscaler{0.4}

\begin{scope}[shift={(0,0)}]

\begin{axis}[                       %axis 1
/pgf/number format/.cd,
        use comma,
        1000 sep={},
    axis lines = box,
    x=\figXscaler*1cm/3,
     y=\figYscaler*7cm/1, 
    axis line style = very thick,
    every axis plot/.append style={very thick},
    xlabel = {$\lambda$ (nm)},
    ylabel = {Transmission},
    xmin = 1500, xmax = 1560, ymin = 0.45, ymax = 2.65,
    ytick={0.6,0.8,...,2.4},
    legend pos=south west,
    yticklabels={,,},
    %ymajorgrids=true,
    %grid style=dashed,
    % width = 8cm,
]

    \addplot[
    color = black]
table[]
{Data/SLR_200K.dat};
\addlegendentry{200 K}
\addplot[
    color = blue]
table[]
{Data/SLR_250K.dat};
\addlegendentry{250 K}
\addplot[
    color = yellow]
table[]
{Data/SLR_300K.dat};
\addlegendentry{300 K}
\addplot[
    color = pink]
table[]
{Data/SLR_350K.dat};
\addlegendentry{350 K}
\addplot[
    color = red]
table[]
{Data/SLR_400K.dat};
\addlegendentry{400 K}

\foreach \z in {0.6,0.8,...,2.4}{
        \addplot[samples=2,
        domain=1450:1590,
        %style = dashed,
        style= thick,
        color = gray,
    ]
        {\z};
    };
\end{axis}
\end{scope}

%%%%%%%%%%%%%%%%%%%%%%%%%%%%
    
\begin{scope}[shift={(8.2,0)}]
\begin{axis}[                           %AXIS3
    axis lines = box,
    /pgf/number format/.cd,
        use comma,
        1000 sep={},
    x=\figXscaler*0.6cm/7,
     y=\figYscaler*1cm/28,
    axis line style = very thick,
    every axis plot/.append style={ultra thick},
    xlabel = {$T$(K)},
    ylabel = {$Q$},
    xmin = 195,xmax = 405 , ymin = 600, ymax = 800,
    legend pos=north west,
    legend cell align={left},
    ytick = {550,600,...,850},
    legend pos=north east,
]

\addplot[
    only marks,
    color = black,
    mark size=3pt,
    mark = star]
table[]
{Data/SLR_gold_Q.dat};

\end{axis}
\end{scope}

\begin{scope}[shift={(8.2,3.1)}]

\begin{axis}[                           %AXIS3
    axis lines = box,
    /pgf/number format/.cd,
        use comma,
        1000 sep={},
    x=\figXscaler*0.6cm/7,
     y=\figYscaler*1cm/1,
    axis line style = very thick,
    every axis plot/.append style={ultra thick},
    ylabel = {Extinction (\%)},
    xticklabel =\empty,
    xmin = 195,xmax = 405 , ymin = 29, ymax = 36,
    legend pos=north west,
    legend pos=north east,
    legend cell align={left},
]
    
    \addplot[
    only marks,
    color = black,
    mark size=3pt,
    mark = star]
table[]
{Data/SLR_gold_strength.dat};

\end{axis}
\end{scope}

\end{tikzpicture}
    \caption{a) The transmission spectra calculated using LSA for homogeneous, temperature independent surroundings for temperature dependent gold NPs. A horizontal line represents a drop of 20\,\% in transmittance. The extinction b) and $Q$-factor c) of the SLR in different temperatures. The spectral location of the peak hardly showed any change with temperature.}
    \label{fig:LSA_gold}
\end{figure}
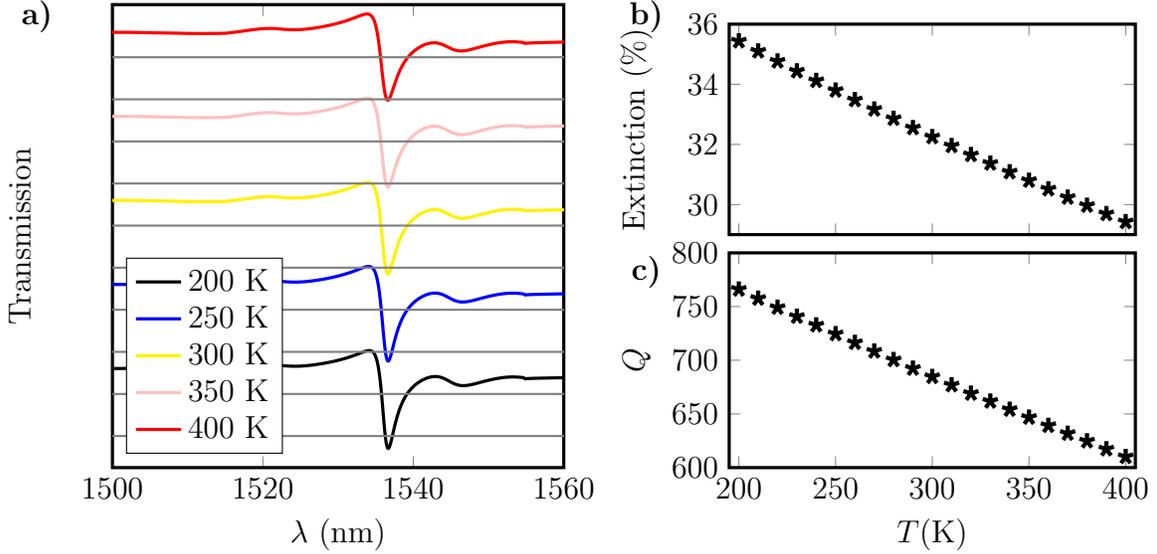

The resulting temperature-dependent permittivity was used in the Equation \eqref{eq:polarizability} for polarizability $\alpha$, which in turn was used in the LSA calculations to estimate how the temperature dependence of the permittivity affects the formation of the SLR. The results are shown in figure \ref{fig:LSA_gold}. Even with a broad temperature range used in the calculations ($\Delta T = 200$~K), the line shapes of the SLRs were almost unnoticeably modified.

It is evident that the slight temperature dependence of the permittivity of the metal has little effect on the SLR in the small temperature range ($\Delta T = 30$~K) we used in our work. Attempts to model the effects of the temperature dependence of aluminum were therefore omitted. However, we note that in the case of perfectly homogeneous environment, the intrinsic temperature dependence of the metal could significant change the formation of the SLR, when broad enough temperature ranges are used. In particular, looking at our results, it is evident that the $Q$-factors of the SLRs could be increased when samples would be cooled to cryogenic temperatures. Other effect affecting SLRs in homogeneous environments is the shifting of the spectral location of the SLR with temperature dependent refractive index of the homogeneous substrate.

\section{Extraction of peak parameters from experimental and LSA data}
\label{SM:LSAfits}
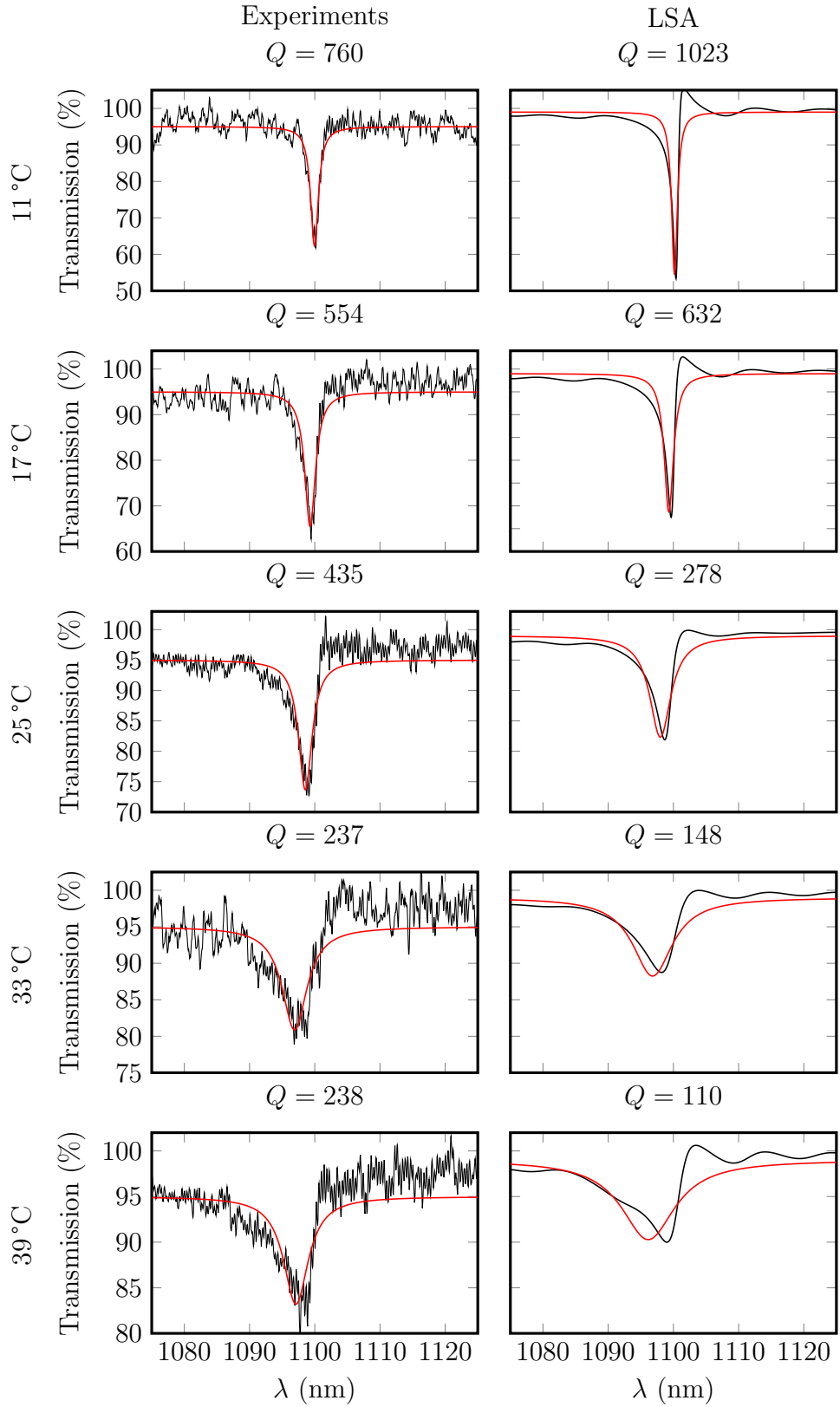
\begin{figure}
    \centering
    \begin{tikzpicture}
%%%% NODES
\node[rotate=90] at (-2,17.5) {\SI{11}{\celsius}};
\node[rotate=90] at (-2,13.5) {\SI{17}{\celsius}};
\node[rotate=90] at (-2,9.5) {\SI{25}{\celsius}};
\node[rotate=90] at (-2,5.5) {\SI{33}{\celsius}};
\node[rotate=90] at (-2,1.5) {\SI{39}{\celsius}};
\node[rotate=0] at (2.5,20.2) {Experiments};
\node[rotate=0] at (8,20.2) {LSA};
\xdef\figXscaler{0.3}
 \xdef\figYscaler{0.4}

\begin{scope}[shift={(0,16)}]                %(0,4)

\begin{axis}[                       %axis 1
/pgf/number format/.cd,
 %       use comma,
        1000 sep={},
    axis lines = box,
    x=\figXscaler*1cm/3,
     y=\figYscaler*7cm/50, 
    axis line style = very thick,
    every axis plot/.append style={very thick},
    title = {$Q = 760$ },
    ylabel = {Transmission~(\%)},
    xmin = 1075, xmax = 1125, ymin = 50, ymax = 105,
    ytick={50,60,...,100},
    legend pos=south west,
    xticklabels={,,},
    %ymajorgrids=true,
    %grid style=dashed,
    % width = 8cm,
]

    \addplot[
    color = black,
    style = thin]
table[]
{Data/Lorentzians/D4_11C.dat};
%\addlegendentry{200 K}
    \addplot[
    color = red,
    style = semithick]
table[]
{Data/Lorentzians/lorentz_fit_11C.dat};
%\addlegendentry{200 K}
\end{axis}
\end{scope}

\begin{scope}[shift={(5.5,16)}]                      %(1,4)

\begin{axis}[                     
/pgf/number format/.cd,
        1000 sep={},
    axis lines = box,
    x=\figXscaler*1cm/3,
     y=\figYscaler*7cm/50, 
    axis line style = very thick,
    every axis plot/.append style={very thick},
     title = {$Q = 1023$ },
    xmin = 1075, xmax = 1125, ymin = 50, ymax = 105,
    ytick={50,60,...,100},
    legend pos=south west,
    xticklabels={,,},
    yticklabels={,,},
    %ymajorgrids=true,
    %grid style=dashed,
    % width = 8cm,
]

    \addplot[
    color = black,
    style = semithick]
table[]
{Data/Lorentzians/LSA_11C.dat};
%\addlegendentry{200 K}
    \addplot[
    color = red,
    style = semithick]
table[]
{Data/Lorentzians/LSA_lorentz_11C.dat};
%\addlegendentry{200 K}
\end{axis}
\end{scope}

\begin{scope}[shift={(0,12)}]                %(0,3)

\begin{axis}[                       %axis 1
/pgf/number format/.cd,
 %       use comma,
        1000 sep={},
    axis lines = box,
    x=\figXscaler*1cm/3,
     y=\figYscaler*7cm/50/0.8, 
    axis line style = very thick,
    every axis plot/.append style={very thick},
    title = {$Q = 554$ },
    ylabel = {Transmission~(\%)},
    xmin = 1075, xmax = 1125, ymin = 60, ymax = 104,
    ytick={50,60,...,100},
    legend pos=south west,
    xticklabels={,,},
    %ymajorgrids=true,
    %grid style=dashed,
    % width = 8cm,
]

    \addplot[
    color = black,
    style = thin]
table[]
{Data/Lorentzians/D4_17C.dat};
%\addlegendentry{200 K}
    \addplot[
    color = red,
    style = semithick]
table[]
{Data/Lorentzians/lorentz_fit_17C.dat};
%\addlegendentry{200 K}
\end{axis}
\end{scope}

\begin{scope}[shift={(5.5,12)}]                      %(1,3)

\begin{axis}[                     
/pgf/number format/.cd,
 %       use comma,
        1000 sep={},
    axis lines = box,
    x=\figXscaler*1cm/3,
     y=\figYscaler*7cm/50/0.8, 
    axis line style = very thick,
    every axis plot/.append style={very thick},
     title = {$Q = 632$ },
    xmin = 1075, xmax = 1125, ymin = 60, ymax = 104,
    ytick={50,55,...,100},
    legend pos=south west,
    xticklabels={,,},
    yticklabels={,,},
    %ymajorgrids=true,
    %grid style=dashed,
    % width = 8cm,
]

    \addplot[
    color = black,
    style = semithick]
table[]
{Data/Lorentzians/LSA_17C.dat};
%\addlegendentry{200 K}
    \addplot[
    color = red,
    style = semithick]
table[]
{Data/Lorentzians/LSA_lorentz_17C.dat};
%\addlegendentry{200 K}
\end{axis}
\end{scope}
%%%%%%%%%%%%%%%%%%%%%%%%%%%%

\begin{scope}[shift={(0,8)}]                %(0,2)

\begin{axis}[                       %axis 2
/pgf/number format/.cd,
 %       use comma,
        1000 sep={},
    axis lines = box,
    x=\figXscaler*1cm/3,
     y=\figYscaler*7cm/50/0.6, 
    axis line style = very thick,
    every axis plot/.append style={very thick},
     title = {$Q = 435$ },
    ylabel = {Transmission~(\%)},
    xmin = 1075, xmax = 1125, ymin = 70, ymax = 103,
    ytick={50,55,...,100},
    legend pos=south west,
    xticklabels={,,},
    %ymajorgrids=true,
    %grid style=dashed,
    % width = 8cm,
]

    \addplot[
    color = black,
    style = thin]
table[]
{Data/Lorentzians/D4_25C.dat};
%\addlegendentry{200 K}
    \addplot[
    color = red,
    style = semithick]
table[]
{Data/Lorentzians/lorentz_fit_25C.dat};
%\addlegendentry{200 K}
\end{axis}
\end{scope}

%%%%%%%%%%%%%%%%%%%%%%%%%%%%

\begin{scope}[shift={(5.5,8)}]                  %(1,2)

\begin{axis}[                       %axis 3
/pgf/number format/.cd,
 %       use comma,
        1000 sep={},
    axis lines = box,
    x=\figXscaler*1cm/3,
     y=\figYscaler*7cm/50/0.6, 
    axis line style = very thick,
    every axis plot/.append style={very thick},
     title = {$Q = 278$ },
    xmin = 1075, xmax = 1125, ymin = 70, ymax = 103,
    ytick={80,85,...,100},
    legend pos=south west,
    yticklabels={,,},
    xticklabels={,,},
    %ymajorgrids=true,
    %grid style=dashed,
    % width = 8cm,
]

    \addplot[
    color = black,
    style = semithick]
table[]
{Data/Lorentzians/LSA_25C.dat};
%\addlegendentry{200 K}
    \addplot[
    color = red,
    style = semithick]
table[]
{Data/Lorentzians/LSA_lorentz_25C.dat};
%\addlegendentry{200 K}
\end{axis}
\end{scope}

\begin{scope}[shift={(5.5,4)}]              %(1,1)

\begin{axis}[                       %axis 3
/pgf/number format/.cd,
 %       use comma,
        1000 sep={},
    axis lines = box,
    x=\figXscaler*1cm/3,
     y=\figYscaler*7cm/50/0.5, 
    axis line style = very thick,
    every axis plot/.append style={very thick},
     title = {$Q = 148$ },
    xmin = 1075, xmax = 1125, ymin = 75, ymax = 102.5,
    ytick={70,75,...,100},
    legend pos=south west,
    yticklabels={,,},
    xticklabels={,,},
    %ymajorgrids=true,
    %grid style=dashed,
    % width = 8cm,
]

    \addplot[
    color = black,
    style = semithick]
table[]
{Data/Lorentzians/LSA_33C.dat};
%\addlegendentry{200 K}
    \addplot[
    color = red,
    style = semithick]
table[]
{Data/Lorentzians/LSA_lorentz_33C.dat};
%\addlegendentry{200 K}
\end{axis}
\end{scope}

\begin{scope}[shift={(0,4)}]                %(0,1)

\begin{axis}[                       %axis 3
/pgf/number format/.cd,
 %       use comma,
        1000 sep={},
    axis lines = box,
    x=\figXscaler*1cm/3,
     y=\figYscaler*7cm/50/0.5, 
    axis line style = very thick,
    every axis plot/.append style={very thick},
     title = {$Q = 237$ },
    ylabel = {Transmission~(\%)},
    xmin = 1075, xmax = 1125, ymin = 75, ymax = 102.5,
    ytick={70,75,...,100},
    legend pos=south west,
    xticklabels={,,},
    %ymajorgrids=true,
    %grid style=dashed,
    % width = 8cm,
]

    \addplot[
    color = black,
    style = thin]
table[]
{Data/Lorentzians/D4_33C.dat};
%\addlegendentry{200 K}
    \addplot[
    color = red,
    style = semithick]
table[]
{Data/Lorentzians/lorentz_fit_33C.dat};
%\addlegendentry{200 K}
\end{axis}
\end{scope}

\begin{scope}[shift={(5.5,0)}]              %(1,0)

\begin{axis}[                       %axis 3
/pgf/number format/.cd,
 %       use comma,
        1000 sep={},
    axis lines = box,
    x=\figXscaler*1cm/3,
     y=\figYscaler*7cm/50/0.4, 
    axis line style = very thick,
    every axis plot/.append style={very thick},
     title = {$Q = 110$ },
    xlabel = {$\lambda$ (nm)},
    xmin = 1075, xmax = 1125, ymin = 80, ymax = 102,
    ytick={80,85,...,100},
    legend pos=south west,
    yticklabels={,,},
    %xticklabels={,,},
    %ymajorgrids=true,
    %grid style=dashed,
    % width = 8cm,
]

    \addplot[
    color = black,
    style = semithick]
table[]
{Data/Lorentzians/LSA_39C.dat};
%\addlegendentry{200 K}
    \addplot[
    color = red,
    style = semithick]
table[]
{Data/Lorentzians/LSA_lorentz_39C.dat};
%\addlegendentry{200 K}
\end{axis}
\end{scope}

\begin{scope}[shift={(0,0)}]                %(0,0)

\begin{axis}[                       %axis 3
/pgf/number format/.cd,
 %       use comma,
        1000 sep={},
    axis lines = box,
    x=\figXscaler*1cm/3,
     y=\figYscaler*7cm/50/0.4, 
    axis line style = very thick,
    every axis plot/.append style={very thick},
     title = {$Q = 238$ },
    xlabel = {$\lambda$ (nm)},
    ylabel = {Transmission~(\%)},
    xmin = 1075, xmax = 1125, ymin = 80, ymax = 102,
    ytick={80,85,...,100},
    legend pos=south west,
    %xticklabels={,,},
    %ymajorgrids=true,
    %grid style=dashed,
    % width = 8cm,
]

    \addplot[
    color = black,
    style = thin]
table[]
{Data/Lorentzians/D4_39C.dat};
%\addlegendentry{200 K}
    \addplot[
    color = red,
    style = semithick]
table[]
{Data/Lorentzians/lorentz_fit_39C.dat};
%\addlegendentry{200 K}
\end{axis}
\end{scope}
\end{tikzpicture}
    \caption{Fitted Lorentzian lineshapes (red) on top of experimental data (black). Take note of the different $y$-axes for different temperatures.}
    \label{fig:fits}
\end{figure}
All $Q$-factors, peak locations $\lambda_c$ and extinctions were determined in this work by fitting a Lorentzian line shape to the extinction spectra. Lorentzian is a line shape given by equation
\begin{equation}
f\left(\lambda ; \lambda_{c}, \gamma_0, I\right)=I\left[\frac{\gamma^{2}}{\left(\lambda-\lambda_{c}\right)^{2}+\gamma^{2}}\right],
\end{equation}
where $\lambda$ is the wavelength, $I$ is the peak extinction of the resonance peak, and $\gamma_0$ is the half width at half maximum. The $Q$-factor is therefore given by 
\begin{equation}
Q = \frac{\lambda_c^{-1}}{(\lambda_c+\gamma_0)^{-1}-(\lambda_c-\gamma_0)^{-1}}.
\end{equation}
The fitted models are shown along with the experimental data on Figure \ref{fig:fits}.

\end{document}